\documentclass[twocolumn,showpacs]{revtex4}
\usepackage{graphicx}% Include figure files
\usepackage{dcolumn}% Align table columns on decimal point
\input epsf
\def\be{\begin{equation}}
\def\ee{\end{equation}}
\def\bea{\begin{eqnarray}}
\def\eea{\end{eqnarray}}

\begin{document}
 \tighten
%\input epsf
%%\draft
\renewcommand{\topfraction}{0.8}
\vskip 3cm

\

\preprint{SU-ITP-03/01,\,  SLAC-PUB-9630, \,TIFR/TH/03-03}
\title {\Large\bf de Sitter Vacua in String Theory }
 \author{\bf Shamit Kachru,$^{1,2}$  Renata Kallosh,$^1$ Andrei Linde$^1$
  and Sandip P.  Trivedi$^3$ }
\address{ {$^1$Department
  of Physics, Stanford University, Stanford, CA 94305-4060,
USA}    }
\address{$^2$SLAC, Stanford University, Stanford, CA 94309, USA}
\address{$^3$TIFR, Homi Bhabha Road, Mumbai 400 005, INDIA}
{\begin{abstract}
We outline the construction of metastable de Sitter vacua of
type IIB string theory.  Our starting point is highly
warped IIB compactifications
with nontrivial NS and RR three-form fluxes.  By incorporating
known corrections to the superpotential from
Euclidean D-brane instantons or gaugino condensation, one
can make models with all moduli
fixed, yielding a supersymmetric AdS vacuum.  Inclusion of a small number of ${\overline{D3}}$ branes in the resulting warped geometry allows one to uplift the AdS minimum and make it a metastable de Sitter ground state.
The lifetime of our metastable de Sitter vacua is 
much greater than the cosmological timescale 
of $10^{10}$ years.  We also 
prove, under 
certain conditions,  that  the 
lifetime of dS 
space in string theory will always be shorter than the recurrence time.

\end{abstract}}
\maketitle
%\tableofcontents{}
\section{Introduction}
There has recently been a great deal of interest in finding de Sitter (dS)
vacua of supergravity and string theory.  This is motivated in part by
the desire to construct possible models for late-time cosmology (since
a small positive cosmological constant seems to be required by recent
data \cite{data}), and in part by more conceptual worries that arise
in the study of dS quantum gravity (as discussed, for instance, in
\cite{conceptual}).  The no-go theorem of \cite{maldanun} guarantees
that such solutions cannot be obtained in string or M-theory
by using only the lowest
order terms in the 10d or 11d supergravity action, but one expects that
 corrections to the leading order Lagrangian in the $g_s$ or
$\alpha^{\prime}$ expansion
 or inclusion of extended sources (branes) should
improve the situation.  Indeed, a careful discussion of how such
additional sources (which are present in string theory) invalidate
the no-go theorem for warped backgrounds and allow one
to find highly warped compactifications appears in \cite{GKP}.
Additional sources which violate the assumptions of the theorem were shown
to yield dS vacua in ${\it noncritical}$ string theory in \cite{MSS}.

Here, we  use our knowledge of  quantum corrections and extended objects
in string theory to argue that there are dS
solutions of ordinary critical
string theory.
Our basic strategy is to first freeze all the moduli present in the
 compactification,
while preserving supersymmetry.
We then
add extra effects that break supersymmetry in a controlled way and lift the
minimum of the potential to a  positive value, yielding dS space.
To illustrate the construction we work  in the
specific context of IIB string theory
compactified on a Calabi-Yau (CY) manifold in the presence of flux.
As described in \cite{GKP} such 
constructions allow one to fix the complex structure moduli,
but not the K\"ahler moduli of the compactification. 
In particular, to leading order in $\alpha^{\prime}$ and $g_s$, the Lagrangian
possesses a
no-scale structure which does not fix the overall volume (we shall
assume that this is the ${\it only}$ K\"ahler modulus
in the rest of this note; it is of course possible to construct
explicit models which
have this property).
In order to
achieve the first
step of fixing all moduli, we therefore need to consider corrections
which violate the no-scale structure. Here  we focus on quantum
non-perturbative corrections to the superpotential
which are calculable and  show that
these can lead to supersymmetry preserving AdS vacua in which
the volume modulus is fixed in a controlled manner.

Having frozen all  moduli we then introduce  supersymmetry breaking
by  adding a few ${\overline{D3}}$ branes in the compactification;
this is a compactified version of the situation discussed in \cite{KPV}.
The addition of ${\overline{D3}}$ branes does not introduce additional
moduli: their worldvolume scalars are frozen by a potential generated
by the background fluxes \cite{KPV}.
Inclusion of anti-D3 branes in the absence of other corrections
yields a run-away to infinite volume of the compact space (since the
energy density in the ${\overline{D3}}$ branes generates a tadpole
for the volume modulus).  However, we show that in the presence of the quantum
corrections we have described, in a sufficiently warped background
the ${\overline{D3}}$
tension can be a small enough correction to lift the formerly
AdS vacuum to positive
cosmological constant, without destabilizing the minimum.

The extent of supersymmetry breaking
and also the resulting cosmological constant
of the dS minimum can be varied in our construction, within a range, in two
ways.  One may vary the
number of  ${\overline{D3}}$ branes which are introduced in the above manner,
and one can also vary the warping in the compactification (by tuning the number of flux quanta
through various cycles).  It is important to note that this corresponds
to a freedom to tune ${\it discrete}$ parameters, so while
fine-tuning is possible, one should not expect to be able
to tune to arbitrarily high precision.
Since there is still a vacuum at very large radius with approximately
zero energy (this is the Dine-Seiberg runaway vacuum \cite{dinesei}),
any dS minimum is only a false vacuum;
but it is only
destabilized by tunneling effects, and we argue that the lifetimes one can
achieve are extremely long.
In addition, we argue that under the assumption that
the potential between the dS minimum and the Dine-Seiberg vacuum at
infinity is positive (which will be true in any simple examples, since
a single dS maximum is the only intervening critical point), the lifetime
of the dS minimum is ${\it always}$ shorter than the timescale for
Poincar\'{e} recurrences discussed in \cite{Susskind}.

While our emphasis in this paper is on dS vacua it is worth remarking that the
first step of our construction, which
freezes all moduli, is of interest in its own right (for another recent
approach to this problem, see \cite{Acharya}; other recent ideas about fixing
the volume modulus, while leaving other moduli unfixed, appear in
\cite{john} and references therein).
In fact moduli stabilization has been an important open
question in string theory and related
phenomenology. We show
that this can be achieved by putting together a few different
effects, all of which are quite well understood by now.
Preserving susy allows us to carry out this analysis with control.
Once all
moduli are frozen,  susy breaking effects other than the introduction of
${\overline{D3}}$ branes alone can also be considered.
Some of these, like the D3/D7 inflationary models of \cite{DasKal}, or
models with
both branes and anti-branes \cite{bantib}, are also
of possible cosmological interest.

A more complete discussion of the possible cosmological
toy models that one can construct using combinations of the ingredients
described in this paper, in the spirit of \cite{KL}, will appear in
\cite{toappear}.

We should note that there has been a great deal of interesting
work on constructing dS solutions in supergravity and string theory.
dS minima of 4d gauged supergravities which do not as yet have a known
string theory embedding appeared in \cite{Fre,Panda}, while
dS compactifications of gauged 6d supergravity appear in \cite{Quevedo}.
The work of \cite{MSS} constructs
dS vacua in supercritical string theory using many of the same
ingredients which arise here.  Finally, the importance of
using fluxes in the cosmological context was stressed in
\cite{BP}, where however the
problem of moduli stabilization was left as a black box (related ideas appeared
in \cite{Feng}).
Cosmology of the simplest flux compactifications, on the $T^6/Z_2$
orientifold \cite{DRS,KST,PolFrey} (whose gauged supergravity
description is worked out in \cite{Ferrara}), was
recently investigated in \cite{Frey}.

\section{Flux compactifications of IIB string theory, including corrections}

In this section, we briefly describe the required knowledge of
flux compactifications of type IIB string theory.  In section IIA we
describe the models of \cite{GKP}, and in section IIB we enumerate
various quantum corrections which can modify the superpotential
and K\"ahler potential used in \cite{GKP}.
In section IIC, we show that incorporating the generic corrections
can yield (supersymmetric) AdS minima with all moduli stabilized.

\subsection{Calabi-Yau orientifolds with flux}

We start with F-theory \cite{Vafa} compactified on an elliptic
CY fourfold $X$.
The F-theory fourfold is a useful way of encoding the data of a solution
of type IIB string theory; the base manifold $M$ of the fibration encodes
the IIB geometry, while the variation of the complex
structure $\tau$ of the elliptic fiber describes
the profile of the IIB axio-dilaton.
In such a model, one has a tadpole condition
\be
{\chi(X)\over 24} ~=~N_{D3} + {1\over {2 \kappa_{10}^2 T_3}}
\int_{M} H_{3} \wedge F_3~.
\label{tadpole}
\ee
Here $T_3$ is the tension of a $D3$ brane, $N_{D3}$ is the
net number of ($D3 - {\overline{D3}}$) branes
one has inserted filling the noncompact dimensions, and $H_3$, $F_3$
are the three-form fluxes in the IIB theory which arise in the
NS and RR sector, respectively.
As shown by Sen \cite{Sen}, in the absence of flux,
it is always possible to deform such
an F-theory model to a locus in moduli space where it can be thought
of as an orientifold of a IIB Calabi-Yau compactification.
For this reason, we will use the language of IIB orientifolds,
with $M$ being the Calabi-Yau threefold which is orientifolded.
In this language, the term ${\chi(X)\over 24}$ counts the negative
D3-brane charge coming from the $O3$ planes and the induced D3
charge on $D7$ branes, while the terms on the right-hand side
count the net D3 charge from transverse branes and fluxes in
the CY manifold.
As in \cite{GKP}, we will assume we are working with a model
having only one K\"ahler modulus, so $h^{1,1}(M) = 1$
($h^{1,1}(X)=2$, and one modulus is frozen in taking the
F-theory limit, where one shrinks the elliptic fiber).  Such models
can be explicitly constructed, by e.g. using the examples
of CY fourfolds in \cite{fourlist} or by explicitly constructing
orientifolds of known CY threefolds with $h^{1,1}=1$.

In the presence of the nonzero fluxes, one generates a superpotential
for the Calabi-Yau moduli, which follows from \cite{GVW} (see also
\cite{Taylor,Klemm}) and is
of the form
\be
W ~=~\int_{M} G_{3} \wedge \Omega
\label{fluxsup}
\ee
where $G_3 = F_3 - \tau H_3$, with $\tau$ the IIB axiodilaton.
Combining this with the tree-level K\"ahler potential
\be
K ~=~-3 \ln[-i(\rho - \bar \rho)] - \ln[-i(\tau - \bar \tau)] -
\ln[-i\int_{M} \Omega \wedge {\overline{\Omega}}]
\label{treekahler}
\ee
where $\rho$ is the single volume modulus ($\rho=b/\sqrt{2}+ i e^{4u -\phi}$;
our conventions are as in
\cite{GKP}),
and using the standard $N=1$ supergravity formula for the potential,
one finds
\bea
V&& ~=~e^{K}\left( \sum_{a,b} g^{a\bar b} D_{a}W \overline {D_b W}
- 3|W|^2 \right)\to \nonumber\\
 &&e^{K}(\sum_{i,j} g^{i \bar j} D_i W
\overline{D_j W})
\label{treepot}
\eea
Here, $a,b$ runs over all moduli fields, while $i,j$ runs over
all moduli fields except $\rho$; and we see that because $\rho$
does not appear in (\ref{fluxsup}), it cancels out of the potential
energy (\ref{treepot}), leaving the positive semi-definite
potential characteristic of no-scale
models \cite{noscale}.

One should use this potential as follows.  Fix an integral choice of
$H_3, F_3$ in $H^{3}(M,{\bf Z})$; then, the potential (\ref{treepot})
fixes the moduli at values where the resulting $G_3$ is imaginary
self-dual (ISD).  Supersymmetric solutions furthermore require $G_3$
to be type (2,1) (more generally, $G_3$ would have a (0,3) piece).
Thus in supersymmetric solutions $W = 0$ on the vacuum, while
in the nonsupersymmetric solutions, $W = W_0$, a constant which is determined
by the (0,3) piece of $G_3$.  In generic solutions, the
complex structure moduli of the F-theory fourfold (in IIB language,
the complex structure moduli, the dilaton, and the moduli of D7
branes) are completely fixed, leaving only the volume modulus
$\rho$.  The scale of the masses $m$ for the moduli which are fixed is
\be
m \sim {\alpha^\prime \over R^3}
\label{masses}
\ee
where $R$ is the radius of the manifold ($\rm Im\,\rho $ scales
like $R^4$).
In this approximation, $R$ is unfixed.  By tuning flux quanta, it is
possible (at least in some cases) to fix $g_s$ at small values, though
not arbitrarily small.

There is one last point we will need to use in section III.
The fluxes (and any transverse branes) serve as sources
for a warp factor.  Therefore, such models with branes and flux
are generically warped compactifications.  In fact, as shown in
\cite{GKP}, following earlier work of \cite{Beckers,Verlinde,DRS,Mayr,Greene},
it is possible to construct models with exponentially large warping.
One can write the Einstein frame metric of the compactification as
\be
ds_{10}^2 ~=~e^{2A(y)} \eta_{\mu\nu}dx^{\mu}dx^{\nu} +
e^{-2A(y)} \tilde g_{mn}(y) dy^m dy^n
\label{metric}
\ee
with $y$ coordinatizing the compact dimensions, and $\tilde g_{mn}$
the unwarped metric on $M$ (so in the orientifold limit, it is a
Calabi-Yau metric).  Then it is shown in \cite{GKP} (by compactifying
the Klebanov-Strassler solution \cite{KS}) that one can
construct models parametrized by flux integers $M,K$ such that
\be
e^{A_{min}} \sim \exp[-(2\pi K)/(3 g_s M)]
\label{expwarp}
\ee
with $e^A$ being of order one at generic points.
This means in particular that with reasonably small flux quanta, one
can generate exponentially large ratios of scales in such models.

In the following, we will assume that $g_s$ and the complex structure
moduli have been fixed at the scale
(\ref{masses}) by a suitable choice of flux, and concentrate on
an effective field theory for the volume modulus $\rho$.  This is
self-consistent, in that the final mass for $\rho$ will be
small compared to (\ref{masses}). One last comment.  Light states could
also arise from modes living in the throat region which experiences large 
warping.  We assume here that any such excitations are gapped, as in \cite{KS}.
Then typically the  $\rho$ modulus, which has a Planck scale
suppressed mass, will be much lighter than these excitations and we can neglect them
as well in the low energy theory.

\subsection{Corrections to the no-scale models}

Here, we write down two  known sources of corrections to the no-scale
models,  both   parametrize possible corrections to the superpotential
(\ref{fluxsup}).
Then, in section IIC,
we show that including either correction to the superpotential
yields supersymmetric models with AdS vacua.

1)  Witten has argued that in type IIB compactifications of this type,
there can be corrections to the superpotential coming from Euclidean
D3 branes \cite{witsup}.  This happens when the fourfold
$X$ used for F-theory compactification admits divisors of arithmetic
genus one, which project to four-cycles in the base $M$.
In the presence of such instantons, there is a correction to the
superpotential which at large volume yields a new term
\be
W_{\rm inst} = T(z_i) \exp(2\pi i \rho)
\label{winst}
\ee
where $T(z_i)$ is a complex structure dependent one-loop determinant,
and the leading exponential dependence comes from the action of a
Euclidean D3 brane wrapping a four-cycle in $M$.
Since the $z_i$ and the dilaton are fixed by the fluxes at a scale
(\ref{masses}), we can integrate them out and view
(\ref{winst}) as simply providing a superpotential for the
volume modulus.

2)  In general models of this sort, one finds (at special
loci in the complex structure moduli space) non-Abelian gauge groups arising
from geometric singularities in $X$, or in type IIB language, from
stacks of D7 branes wrapping 4-cycles in $M$.
Assume that the fluxes have fixed one at a point in moduli space
where this phenomenon occurs (examples appear in \cite{santrip},
for instance).
Consider
a stack of $N_c$ coincident branes.  The 4d gauge coupling
of the $SU(N_c)$ Yang-Mills theory on such wrapped branes (we ignore
the decoupled $U(1)$ factor) satisfies
\be
{8 \pi^2\over g_{YM}^2} = 2 \pi {R^4 \over g_s}= 2 \pi \, {\rm Im \, \rho}~.
\label{sevgauge}
\ee
Since the complex structure moduli of $X$ are completely fixed,
the D7 brane moduli (at least in cases where the 4-cycle being
wrapped has vanishing $h^1$, which are easy to arrange) are
also fixed.  Therefore, any charged matter fields (which would
create a Higgs branch for the D7 gauge theory)
have also been given a mass at a high-scale; and the low-energy
theory is pure $N=1$ supersymmetric $SU(N_c)$ gauge theory.
This theory undergoes gluino condensation, which results in a
nonperturbative superpotential
\be
W_{\rm gauge} =\Lambda_{N_c}^3 = A e^{2\pi i  \rho \over N_c}
\label{gaugino}
\ee
where $\Lambda_{N_c}$ is the dynamical scale of the gauge theory,
and the coefficient $A$ is determined by the energy  scale below which the
the SQCD theory is valid (There  are also threshold corrections
in general, these contribute subleading effects.)
We see that this leads to an exponential superpotential for
$\rho$ similar to the one above (but with a fractional multiple
of $\rho$ in the exponent, since the gaugino condensate looks
like a fractional instanton effect in $W$).

So effects 1) and 2) have rather similar consequences for
our analysis; we will simply assume that there is an exponential
superpotential for $\rho$ at large volume.
In our companion paper \cite{toappear}, we investigate some
interesting possibilities for cosmology if there are multiple
non-Abelian gauge factors.  Using the fourfolds in \cite{fourlist},
it is easy to construct examples (with $h^{1,1}(X)=2$)
which could yield gauge groups of
total rank up to $\sim 30$.  The results of \cite{kaplouis} suggest
that much larger ranks should be possible.

One important comment is in order before we proceed.
Besides corrections to the superpotential of the kind discussed above,
 there are  also corrections to the K\"ahler potential
(see e.g. \cite{bbhl} for a calculation of some leading corrections). In
our analysis we will ensure
that the volume
modulus is stabilized
at values which are parametrically large compared to the string scale.
This makes our neglect of K\"ahler  corrections self consistent.

\subsection{Supersymmetric AdS Vacua}

Here, we show that the corrections to the superpotential considered above
can stabilize the volume modulus, leading to a susy preserving AdS minimum.
We  perform an analysis of the vacuum structure
just keeping the tree-level K\"ahler
potential
\be
K = - 3 \ln[-i(\rho - \overline{\rho})]
\ee
and a  superpotential
\be
W = W_0 + Ae^{ia\rho}~.
\label{adssup}
\ee
$W_0$ is a tree level contribution which arises from the fluxes. The exponential term
arises from either of the two sources above, and the coefficient $a$ can be
determined accordingly.
In keeping with the fact that the complex structure
moduli and the dilaton have received a mass (\ref{masses}), we
have set them equal to their VEVs and consider only the low-energy theory
of the volume modulus.  To avoid the need to worry about additional
open-string moduli, we assume the tadpole condition (\ref{tadpole}) has
been solved by turning on only flux, i.e. with no additional $D3$ branes.

At a  supersymmetric vacuum  $D_\rho W=0$. We simplify things by setting
the axion in the $\rho$ modulus to zero, and
letting $\rho = i\sigma$. In addition we take  $A,a$ and $W_0$ to be  all {\it real} and $W_0$ negative.
The minimum then lies at
\be
DW=0 \quad \rightarrow \quad   W_0=- Ae^{-a\,\sigma_{cr}}(1+ {2\over 3}\, a\, \sigma_{cr})
\label{susymin}
\ee

The potential, $V= e^{K}\left(  G^{\rho\overline\rho} D_{\rho}W \overline {D_{\rho} W}
- 3|W|^2 \right)$,
  at the minimum is  negative and equal to
\be
V_{\rm AdS}=(-3 e^{K}W^2)_{AdS}= -{a^2 A^2 e^{-2\,a\,\sigma_{cr}}\over 6 \,\sigma_{cr}}
\ee
We see that we have stabilized the volume modulus while preserving supersymmetry.
It is important to note that the AdS minimum is quite generic. Any corrections to the
K\"ahler potential will still result in  a susy minimum which solves (\ref{susymin}).

A few comments are in order before we proceed.
A controlled calculation requires that $\sigma \gg 1$, this ensures that  the
supergravity approximation is valid and the  $\alpha^{\prime}$
corrections to the
K\"ahler potential are under control. It also requires that $a\, \sigma >1$ so that the
contribution to the superpotential from a single (fractional) instanton is reliable.
Generically, if the fluxes break supersymmetry, $W_0 \sim O(1)$, and these conditions will not be
met. However  it is reasonable to expect that by tuning fluxes one can arrange so that
 $W_0 \ll 1$. In these circumstances
 we see from (\ref{susymin}) that $a\, \sigma >1$. Taking $a < 1$, one can then   ensure that
$\sigma \gg 1$, as required.

As an illustrative example we consider $W_0 =- 10^{-4}$, $A=1$, $a =0.1$.
This results in a minimum at $\sigma_{cr}\sim 113$.

\begin{figure}[h!]
\centering\leavevmode\epsfysize=5cm \epsfbox{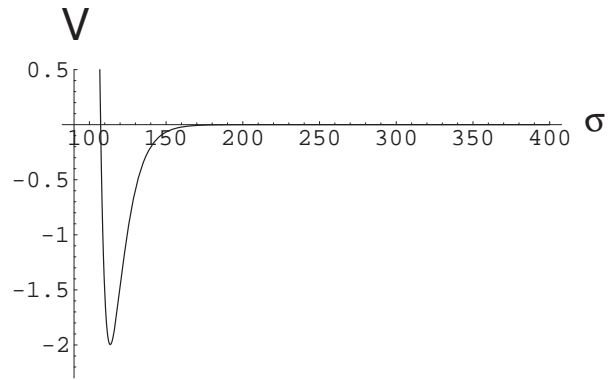}
\caption[fig1]
{Potential (multiplied by $10^{15}$) for the case of
exponential superpotential with $W_0 =- 10^{-4}$, $A=1$, $a =0.1$.
There is an AdS minimum.}
\label{ExpAdS3}
\end{figure}

Another possibility to get a minimum at large volume is to consider
a situation where the fluxes preserve susy, and the  superpotential involves
multiple  exponential terms, i.e. ``racetrack potentials" for the stabilization of $\rho$
\cite{racetrack}.
Such a superpotential could arise from multiple stacks of seven branes wrapping
four cycles which cannot be deformed into each other in a susy preserving manner \cite{toappear}.
In this case by tuning the ranks of the gauge groups appropriately
 one can obtain a parametrically large value of $\sigma$ at the minimum.

Finally it is plausible
that
in some cases, there could be some
other moduli which
have not yet been fixed in this approximation (if, for instance,
the fluxes were nongeneric and left some of the complex structure moduli of the
Calabi-Yau manifold unfixed). Since the dilaton is already fixed by fluxes
and the radial modulus is fixed due to
non-perturbative exponential terms in the superpotential, one might
find interesting and realistic cosmology driven
by these other scalar fields from the compactified string theory.
Here, however, we will neglect this possibility, and focus instead on
another mechanism for uplifting the AdS vacuum to a dS vacuum.

\section{Constructing dS vacua}

In this section, we uplift the supersymmetric AdS vacua of section IIC
to yield dS vacua of string theory.  In section IIIA, we describe the
new ingredient: ${\overline{D3}}$ branes transverse to $M$.  In
section IIIB, we show that for reasonable choices of parameters, the
inclusion of ${\overline{D3}}$ branes yields dS vacua.

\subsection{$\overline{D3}$ branes in ISD fluxes}

In the tadpole condition (\ref{tadpole}), there is a contribution
from both localized $D3$ branes and from fluxes.
To find AdS vacua with no moduli in the previous section, we assumed
that the condition was saturated by turning on fluxes in the compact
manifold.  Now, we assume that in fact we turn on ${\it too~ much}$ flux,
so that (\ref{tadpole}) can only be satisfied by introducing one
${\overline{D3}}$ brane.  In the flux background determining a
solution of the sort described in section II, the $\overline{D3}$
does not have translational moduli; they are fixed by the ISD fluxes,
which generate a potential for the worldvolume scalars.
This kind of situation was studied in
\cite{KPV}.

Now, the tadpole is cancelled, but there is an extra bit of energy
density from the ``extra'' flux and ${\overline {D3}}$ brane.  In fact,
as in equation (73) of \cite{KPV}, one finds that the ${\overline{D3}}$ brane
adds an additional energy:
\be
\delta V = 2 {a_0^4 T_{3} \over  g_s^4} { 1 \over ({\rm Im\, \rho})^3 }
\label{smallen}
\ee
with $a_0$ the warp factor at the location of the ${\overline{D3}}$ brane.
As described in \cite{KPV}, in the presence of ISD fluxes of the sort
characterizing the Klebanov-Strassler throat, any anti-D3 branes
are driven to the end of the throat, where the warp factor is minimized.
It follows from (\ref{expwarp}) that the value of $a_0$ is exponentially
small, and hence in suitable models the inclusion of a ${\overline{D3}}$
adds to the potential an exponentially suppressed term.
The prefactor of $2$ in (\ref{smallen}) arises because the ambient
five-form flux adds a repulsive energy equal to the tension for an ${\overline {D3}}$ brane,
see \cite{maldnast} (above equation (3.9)). Note
also that we are considering solutions
which meet the ISD condition, even in the
presence
of the additional flux required to insert the ${\overline {D3}}$ brane.
There are  corrections to (\ref{smallen}) but these   are quadratic in the number of
${\overline{D3}}$ that are added and are small.

The important point is that due to the warping the addition of the ${\overline {D3}}$ brane
breaks supersymmetry by a very small amount.  In general terms,
we get a term in the potential which goes like
\be
\delta V = { 8 D\over (\rm Im \,\rho)^3}
\label{uplift}
\ee
(the factor of ${\rm 8}$ is added for later convenience).
The coefficient $D$ depends on the number of ${\overline {D3}}$ branes and on the warp
factor at the
end of the throat. These parameters can be altered  by discretely changing the total flux, and the
fluxes which enter in (\ref{expwarp}), respectively. This allows us to vary the coefficient $D$
and the susy breaking in the system, while still keeping them small (More properly, since the flux can only be discretely tuned,
$D$ can
be varied but not with arbitrary precision).  We will see that by tuning
the choice of $D$ one can perturb the AdS vacua of IIC to
produce dS vacua with  a tunable  cosmological constant.
The vacua  will clearly only be metastable, since all of the sources of
energy we have introduced vanish as $\rm Im \,\rho \to \infty$.

\subsection{Uplifting AdS vacua to dS vacua}

We now
add to the potential a term of the form $D/\sigma^3$, as explained above.
For suitable choices of $D$, the AdS minimum will become a dS minimum, but
the rest of the potential does not change too much. There is 
one new important feature, however: there is a dS maximum separating 
the dS
minimum from the vanishing potential at infinity. The potential is:

$${V ~=~ {aAe^{-a\sigma}\over {2\sigma^2}} \left( {1\over 3}\sigma
aA e^{-a\sigma} +  W_0 +  A e^{-a\sigma}\right)}+{D\over \sigma^3}$$
By fine-tuning $D$,
it is easy to have the dS minimum very close to zero.
For the model  $W_0 =- 10^{-4}$, $A=1$, $a =0.1$ $D=3 \times 10^{-9}$
we find the potential (multiplied by $10^{15}$):

\begin{figure}[h!]
\centering\leavevmode\epsfysize=5cm \epsfbox{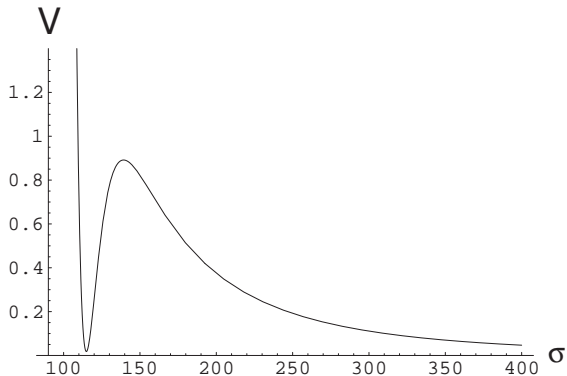}
\caption[fig1]
{Potential (multiplied by $10^{15}$) for the case of
exponential superpotential
and including a $D\over \sigma^3$ correction with $D=3 \times 10^{-9}$ which
uplifts the AdS
minimum to a dS  minimum.}
\label{ExpAdSdS3}
\end{figure}

Note, if one does not require the minimum to be so close to zero, $D$ does not have to be
fine-tuned so precisely. A dS minimum is obtained as long as $D$ lies within a range,
eventually disappearing for large enough D.
If one does fine tune to get the minimum very close to zero, the resulting
potentials are quite steep around the dS minimum.  In this circumstance, 
the new term 
basically uplifts the potential without changing the shape too much around
the minimum, so the $\rho$ field acquires
a surprisingly large mass (relative to the final value of the 
cosmological constant).

It is important to mention that the value of the volume modulus shifts only slightly
in going from the AdS minimum to the new dS minimum. This 
means if the volume was large 
in the AdS minimum to begin with,
 it will continue to be large in the new dS minimum, guaranteeing that
our approximations are valid.

If one wants to use this potential to describe the present stage of acceleration of the universe, one needs to fine-tune the value of the potential in dS minimum to be $V_0 \sim 10^{-120}$ in units of Planck density. In principle, one could achieve it, e.g., by fine tuning $D$. However,
the tuning we can really do by varying the fluxes etc. in the microscopic
string theory is limited, though it may be possible to tune quite well
if there are enough three-cycles in $M$.

\section{How stable is the dS vacuum?}

The radial modulus  $\sigma= \rm Im \, \rho$ has 
a kinetic term ${3\over 4 \sigma^2}(\partial \sigma )^2$ which 
follows from the K\"ahler potential  (\ref{treekahler}). For cosmological purposes it is convenient to switch to the 
canonical variable $\varphi =\sqrt {3\over 2}\ln \sigma= \sqrt {3\over 2}\ln (\rm Im \, \rho)$,
which has a kinetic term ${1\over 2 }(\partial \varphi )^2$. In 
what follows we will use the field $\varphi$ and 
it should not be confused with the dilaton,  $\phi$.

\subsection{General theory}

The dS vacuum 
state $\varphi_0$ 
corresponding to the local minimum of the potential with $V_0>0$ is metastable. Therefore it may decay, and then the universe will roll 
towards large values of the field $\varphi$ and decompactify. Here 
we would like to address two important questions:

1) Do our dS vacua survive for a large number of
Planck times?  For instance, if we fine tune to get a small
cosmological constant, is the dS vacuum sufficiently stable to survive 
during the $10^{10}$ years of the cosmological evolution? If 
the answer is positive, one 
can use the dS minimum for the phenomenological 
description of the current 
stage of acceleration (late-time inflation) of the universe.

2) Is the typical decay time of the dS vacuum longer or shorter 
than the recurrence time $t_r \sim e^{\bf S_0}$, where 
${\bf S_0} = {24\pi^2\over V_0}$ is 
the dS entropy \cite{Gibbons:1976ue}? 
If the  decay time is longer than $t_r \sim e^{\bf S_0}$, one may 
need to address the issues about the  
consistency of the stringy description 
of dS space raised in \cite{conceptual,MSS,Susskind}.

We will argue that the 
lifetime of the dS vacuum in our models is not too short and not too long: it is extremely large in Planck times
(in particular, one can easily make models which live longer than the cosmological timescale $\sim 10^{10}$ years), and it is 
much shorter than the recurrence time $t_r \sim e^{\bf S_0}$.

In order to analyse 
this issue we will
remember, following Coleman and De Luccia   \cite{Coleman:1980aw}, basic 
features of the tunneling theory taking into account gravitational effects.

To describe tunneling from a local minimum at $\varphi = \varphi_0$ one should consider an $O(4)$-invariant Euclidean spacetime with the
metric
\begin{equation}\label{metric2}
ds^2 =d\tau^2 +b^2(\tau)(d \psi^2+ \sin^2 \psi \, d \Omega_2^2) \ .
\end{equation}
The scalar field $\varphi$ and the Euclidean scale factor (three-sphere radius) $b(\tau)$ obey the
equations of motion
\begin{equation}\label{equations}
\varphi''+3{b'\over b}\varphi'=V_{,\varphi},~~~~~ b''= -{b\over 3}  (\varphi'^2 +V) \ ,
 \end{equation}
where primes denote derivatives with respect to $\tau$. (We use the system of units $M_p = 1$.)

These equations have several instanton solutions $(\varphi(\tau),\, b(\tau))$.
The simplest of
them are the $O(5)$ invariant four-spheres one obtains when the field
$\varphi$ sits at one of the extrema of its potential, and $b(\tau)
= H^{-1} \sin H\tau$. Here $H^2 = {V\over 3}$, and $V(\varphi)$ corresponds to one of the extrema.  In our case, there are two trivial solutions of this type. One of them describes time-independent field corresponding to the minimum of the effective potential at $\varphi = \varphi_0$, with $V_0=V(\varphi_0)$. Another one is related to the maximum of the potential at $\varphi =\varphi_1$, with $V_1 = V(\varphi_1)$.

Coleman-De Luccia (CDL) instantons are more complicated.  They describe the field $\varphi(\tau)$ beginning in a vicinity of the false vacuum
$\varphi_0$ at $\tau =0$,  and reaching some constant value $\varphi_{f}> \varphi_1$ at $\tau = \tau_{f}$, where $b(\tau_f)= 0$. It is tempting to interpret CDL instantons as the tunneling trajectories interpolating between the different vacua of the theory. However, one should be careful with this interpretation because the trajectories $\varphi(\tau)$ for CDL instantons {do not} begin exactly in the 
metastable minimum $\varphi_0$ and {do not} end exactly in the absolute minimum of the effective potential. We will discuss this issue later.

According to \cite{Coleman:1980aw}, the tunneling probability is given by
\begin{equation}\label{prob}
P(\varphi) = e^{-S(\varphi)+S_0},
\end{equation}
where $S(\varphi)$ is the Euclidean action  for the 
tunneling trajectory $(\varphi(\tau),\, b(\tau))$, and   $S_0=S(\varphi_0)$ is the Euclidean action for the initial configuration $\varphi = \varphi_0$.  Note that  $S(\varphi)$ in Eq. (\ref{prob}) for the tunneling probability is the integral over the whole instanton solution, rather that the integral over its half  providing the tunneling amplitude.

The tunneling action is given by
\begin{equation}\label{fullaction}
S(\varphi)= \int d^4x
\sqrt{g}\left(-{1\over 2}R+{1\over 2} (\partial \varphi)^2 +V(\varphi)\right).
\end{equation}
In $d=4$ the trace of the Einstein  equation is
$R=(\partial \varphi)^2+4V(\varphi)$. Therefore the total action can be represented by an integral of $V(\varphi)$:
\begin{equation}\label{dsaction}
S(\varphi) = -\int d^4x
\sqrt{g} V(\varphi) = - {2\pi^2\int_0^{\tau_f} d\tau\, b^3(\tau)\, V(\varphi(\tau))}\ .
\end{equation}

The Euclidean action calculated for the false vacuum dS solution $\varphi=\varphi_0$ is given by
\begin{equation}\label{action}
S_0 = - {24\pi^2\over V_0} <0\ .
\end{equation}
Similarly, for the dS maximum  $\varphi =\varphi_1$ one has $S_1 = - {24\pi^2\over V_1}$.

This action for dS space $S_0$ has a simple sign-reversal relation to the entropy of de Sitter space ${\bf S_0}$:
\begin{equation}\label{action2}
{\bf S_0} = - S_0 =+ {24\pi^2\over V_0}\ .
\end{equation}

Therefore the decay time of the metastable dS vacuum $t_{\rm decay} \sim P^{-1}(\varphi)$ can be represented in the following way:
\begin{equation}\label{decaytime}
t_{\rm decay} = e^{S(\varphi)+\bf S_0} = t_r \ e^{S(\varphi)}\ .
\end{equation}
The  semiclassical approximation is applicable only for $|S(\varphi)| \gg 1$.  Eq. (\ref{dsaction}) implies  that for the tunneling through the barrier with $V(\varphi)>0$ (which is the case for the tunneling from dS space to Minkowski space in our model) the action $S(\varphi)$ is always negative, $S(\varphi) < 0$. This means that {\it the decay time of dS space to Minkowski space is exponentially smaller than the 
recurrence time $t_r$}. The existence of the runaway  
vacuum at $\infty$ in field space with zero energy is a standard feature of all string theories \cite{dinesei}.  We 
conclude that the problems
related to the decay time exceeding the recurrence time $t_r$ \cite{conceptual,MSS,Susskind} will not appear in the 
simplest string theory models, where a single dS maximum will separate the dS minimum from $\infty$.   

Now that we found the upper bound on the tunneling time, we will try to estimate the tunneling time in our model using 
some particular instanton solutions. In general, it is 
very difficult to find analytical solutions for the CDL instantons and 
calculate the tunneling probability. We will investigate this problem using two different approaches, which are, in 
a certain sense, opposite to each other: the thin-wall approximation and the no-wall approximation.

\subsection{Thin-wall approximation}

Let us assume that one can describe the present stage of the acceleration of the universe by our model, so 
that  $V_0 \sim 10^{-120}$ in Planck units (more generally, the analysis of this section will work for $V_0$ very small compared to the
starting AdS cosmological constant, which can be arranged by tuning). 
This is a hundred orders of magnitude smaller than the height of the barrier $V_1$ in our model. In Minkowski space, the condition $V_0\ll V_1$ usually means that the thin-wall approximation is applicable. Let us check whether one can use it in our  case. 
 
 In application to our 
scenario (tunneling from dS space with vacuum energy $V_0>0$ to Minkowski space) one can 
represent the results of \cite{Coleman:1980aw} in the following useful form:
\begin{equation}\label{thin}
P = \exp\left({-{{\bf S}(\varphi_0) \over (1 + ( 4V_0 /3T^2) )^{2}}}\right)
\ .
\end{equation}
Here $T$ is not temperature but tension of the bubble wall; in our case
\begin{equation}\label{thin1}
T = \int_{\varphi_0}^\infty d\varphi\, \sqrt{2V (\varphi)}
\ .
\end{equation}
Eq. (\ref{thin}) confirms our general conclusion that the suppression of the tunneling is always smaller than $e^{-{{\bf S}(\varphi_0)}}$, and the decay time is shorter than $t_r$.

There are two limiting cases of special interest: $V_0 \gg T^2$ and $V_0 \ll T^2$.  The meaning of these two conditions becomes clear if we restore $M_p$ in these inequalities: $V_0 M_p^2 \gg T^2$ and $V_0 M_p^2\ll T^2$. If we turn off the gravitational interactions ($M_p \to \infty$), one has $V_0 M_p^2 \gg T^2$, and we obtain the well known result for the tunneling in the thin-wall approximation in Minkowski space: 
\begin{equation}\label{thinnogravity}
P = \exp\left({-{27\pi^2 T^4 \over 2V_0^3}}\right)
\ .
\end{equation} 
From Eq. (\ref{thin1}) one finds that $T\sim \Delta \varphi \sqrt{V_1}$, where $\Delta\varphi$ is the typical width of dS maximum. This means that $V_0 \gg T^2$ (in units $M_p = 1$), and one can ignore the gravitational effects in the thin-wall approximation only if
\begin{equation}\label{thinnogravity2}
{\Delta\phi}\ll \sqrt{V_0\over V_1}
\ .
\end{equation} 

One can easily check that for our model this condition is not satisfied, so we must take gravitational effects into account and study an opposite limit $V_0 \ll T^2$. Then in the first approximation one simply has
\begin{equation}\label{thin2}
P\approx \exp\left({-{{\bf S}(\varphi_0) }}\right)=\exp\left(-{24\pi^2\over V_0}\right) \sim \exp\left(-{10^{122}}\right)
\ .
\end{equation}
Thus, for all practical purposes our dS vacuum is completely stable.

On the other hand, if one expands Eq. (\ref{thin}) in powers of $V_0/T^2$, one finds
 \begin{equation}\label{thin3}
P\approx \exp\left({-{{\bf S}(\varphi_0) }}\right)\times \exp\left({64\pi^2\over T^2}\right)
\ .
\end{equation}
For the sub-Planckian tension $T\ll 1$ one finds that the probability of the tunneling is exponentially larger than $P\approx \exp\left({-{{\bf S}(\varphi_0) }}\right)$. The decay time of dS space due to CDL instantons $t^{\rm CDL}_{\rm decay}$ is much smaller than the recurrence time $t_r \sim \exp\left({-{{\bf S}(\varphi_0) }}\right)$, in agreement with our general result:
 \begin{equation}\label{thin4}
t^{\rm CDL}_{\rm decay}\sim t_r    \exp\left(-{64\pi^2\over T^2}\right)
\ .
\end{equation}

In the thin-wall approximation (for $V_0 \ll T^2$) the radius of the bubble is $R_B =4/T$ \cite{Coleman:1980aw}, whereas the thickness of the wall is approximately $\Delta R_B = T/V_1$. This means that the thin-wall approximation taking gravity into account is valid if  $\Delta R_B\ll R_B$, i.e. $T^2\ll V_1$.

\subsection{``No-wall'' approximation: Hawking-Moss instanton and stochastic approach to tunneling}

Now we will return to the simplest instanton solutions, sitting at the dS minimum $\varphi = \varphi_0$ and the dS maximum
$\varphi =\varphi_1$. According to Hawking and Moss (HM) \cite{Hawking:1981fz}, they describe tunneling through the barrier, with the probability of tunneling suppressed as
 \begin{equation}\label{HM}
P = e^{-S(\varphi_1)+S(\varphi_0))}= \exp\left(-{24\pi^2\over V_0}+{24\pi^2\over V_1}\right)
\ .
\end{equation}
This result may seem  rather controversial because the instanton solution $\varphi(\tau) =\varphi_1$ does not interpolate between the stable vacuum and the false vacuum.

In fact, as we already mentioned, the last problem appears for all CDL solutions as well. These solutions never begin exactly in the false vacuum, so how could they be considered interpolating solutions? This problem disappears in the thin-wall limit, but it shows up very clearly in numerical calculations going beyond the thin wall approximation.

There are several different ways in which one can address this issue. In \cite{GL} it was noticed that instead of considering an exactly constant solution $\varphi = \varphi_1$ one may consider a configuration that coincides with this solution everywhere except a small vicinity of the endpoint at $\tau_f = \pi H^{-1}$. The action involves integration  $\int d\tau\, b^3(\tau) L(g_{\mu\nu},\varphi)$. Since  the scale factor $b(\tau)
= H^{-1} \sin H\tau$ vanishes at  $\tau_f = \pi H^{-1}$, one can make very strong modifications of the solution $\varphi = \varphi_1$ in a small vicinity of $\tau_f$  without changing the action in a significant way. If the action changes by less than $O(1)$, then each such configuration can be used for the description of tunneling despite the fact that it is not a true solution of classical equations of motion near $\tau_f$.
This allows one to bend the solution $\varphi = \varphi_1$ (or the CDL solution) in such a way that it begins to interpolate between the different vacua.

One may also patch HM instantons and CDL instantons to each other in
several different ways, which may provide their alternative
interpretation and justify the results obtained by using CDL and HM
instantons, see e.g. \cite{Linde:1998gs,Bousso:1998vz,Bousso:1998ed}.
More recently a possible interpretation of CDL and  HM instantons  was
suggested in  \cite{Banks:2002nm}. Significant progress was achieved by
Gen and Sasaki in the development of a consistent Hamiltonian approach
to tunneling with gravity \cite{Gen:1999gi}.

But these methods do  not address another problem with the Hawking-Moss tunneling: Since the instanton $\varphi = \varphi_1$ is exactly homogeneous, it seems 
to describe a homogeneous tunneling in the whole universe, which is impossible. One can 
circumvent this problem by claiming that the whole 
universe is reduced to the interior of a single 
causal patch of size $H^{-1}$. However, this approach, which 
is often used for the description of eternal dS space (which does not decay), may be 
less useful in applications to inflationary cosmology since it loses 
information about the whole universe except for a small part of size $H^{-1}$.

The most intuitively transparent description of the (nearly) homogeneous tunneling was provided in \cite{Starobinsky:fx,book,Hardart,LLM} in the context 
of the stochastic approach to inflation. 

One may consider quantum fluctuations of a light scalar field
$\varphi$ with $m^2 = V'' \ll H^2 = V/3$. During 
each time interval $\delta t = H^{-1}$ this 
scalar field experiences quantum jumps with the wavelength $\sim H^{-1}$ and with a typical amplitude $\delta\varphi = H/2\pi$. Then the wavelength of these fluctuations grows exponentially. As a result, quantum fluctuations lead to a local change of the amplitude of the field $\varphi$ which looks homogeneous on the horizon scale $H^{-1}$. From the point of view of a local observer, this process looks like a Brownian motion of the homogeneous scalar field. If the potential has a dS minimum at $\varphi_0$ with $m\ll H$, then eventually the probability distribution to find the  field with the value $\varphi$ becomes time-independent:
\cite{Starobinsky:fx,book,Hardart,LLM}
\begin{equation}\label{E38a}
P(\varphi,\varphi_0) \sim  
\exp\left({24\pi^2\over V(\varphi)}\right)\cdot\, \exp\left(-{24\pi^2\over 
V(\varphi_0)}\right) \ .
\end{equation}

This result was obtained without any 
considerations based on the Euclidean approach to quantum gravity.
It provides a
simple interpretation of the Hawking-Moss tunneling. During inflation, long
wavelength perturbations of the scalar field freeze on top of each other and
form complicated configurations, which, however, look almost homogeneous on the
horizon scale $H^{-1}$. If originally the whole universe was in a state
$\varphi_0$,
the scalar field starts wandering around, and
eventually it reaches the local maximum of the effective potential at $\varphi =
\varphi_1$. According to (\ref{E38a}), the probability of this event is
suppressed by $\exp\left(-{24\pi^2\over V_0}+{24\pi^2\over V_1}\right)$. As soon as the
field $\varphi$ reaches the top of the effective potential, it may fall down to
another minimum, because it looks nearly homogeneous on a scale of horizon, and
gradients of the field $\varphi$ are not strong enough to pull it back to
$\varphi_0$. The probability of this process is given by the Hawking-Moss expression (\ref{HM}). However, this is  not a homogeneous tunneling all over the universe, but rather a  Brownian motion, which  looks
homogeneous on the scale $H^{-1}$ \cite{book}. 

One can take a logarithm of the probability distribution (\ref{E38a}) and find that the entropy of the state with the field stochastically moving in the potential $V(\varphi)$ is given by 
\begin{equation}\label{action2a}
{\bf S(\varphi)} = {24\pi^2\over V(\varphi)}\ .
\end{equation}
This is the simplest derivation of dS entropy that does not involve ambiguous Euclidean calculations. It is valid even for the states outside the dS minimum, as long as the condition $|m^2|\ll H^2$ (i.e. $|V''| \ll V$) remains valid.

This result allows one to obtain a simple interpretation of the HM tunneling, proposed in \cite{Linde:1998gs}. Indeed, Eq. (\ref{HM}) has the standard thermodynamic form
describing the probability of thermal fluctuations
 \begin{equation}\label{HM2}
P = e^{\bf \Delta S}= e^{\bf S(\varphi_1)-\bf S(\varphi_0)}
\ .
\end{equation} 
A similar thermodynamic approach was recently developed in a series of papers by Susskind {\it et al} \cite{Susskind}. 

Since the entropy $S(\varphi_1)= 24\pi^2/V_1$ is positive, one finds another confirmation of our general result: 
\begin{equation}\label{decaytimeHM}
t^{\rm HM}_{\rm decay} = e^{-\bf S_1+\bf S_0} = t_r \ \exp\left(-{24\pi^2\over V_1}\right) \ll t_r\ .
\end{equation}
On the other hand, in our case $S_1\ll S_0$, so in the first approximation one finds, as before, that $t^{\rm HM}_{\rm decay} \approx   t_r \sim e^{10^{122}}$. 

Note 
that the quasi-homogeneous HM tunneling and the CDL tunneling correspond to two different processes; depending on the potential, one of these processes may happen much faster than another. Let us compare the rate of the quasi-homogeneous HM tunneling  with the rate of tunneling due to CDL instantons in the thin wall approximation for $T\ll 1$:
\begin{equation}\label{decaytimeHM2}
{t^{\rm HM}_{\rm decay}\over t^{\rm CDL}_{\rm decay}}= \exp\left[8\pi^2\left({8\over T^2}-{3\over V_1}\right)\right]\ .
\end{equation}
 This shows that $t^{\rm HM}_{\rm decay} < t^{\rm CDL}_{\rm decay}$ (HM tunneling dominates) for $3T^2 > 8V_1$. Meanwhile, in the opposite case, $3T^2 < 8V_1$, the tunneling occurs due to CDL instantons. This is consistent with our estimate of validity of the thin wall approximation, $T^2\ll V_1$. 

It is useful to represent this result in a different way. Let us use the estimate $T\sim \Delta \varphi \sqrt{V_1}$, where $\Delta\varphi$ is the typical width of dS maximum. This estimate implies that $3T^2 > 8V_1$ and $t^{\rm HM}_{\rm decay} < t^{\rm CDL}_{\rm decay}$ if $\Delta\varphi > 1$, i.e. if the width of the maximum is much greater than the Planck mass. For the potentials with the nearly Minkowski minima this condition coincides with the inflationary slow-roll condition $|V''| < V$.

Thus we are coming to the conclusion that the HM tunneling, and the thermodynamic approach discussed above, are most
efficient for the description of the tunneling in the inflationary universe, where their validity has been 
firmly established by the stochastic approach to inflation.
On the other hand, in the situations where the potentials are very thin, $\Delta\phi \ll 1$, one should use the 
CDL instantons, and the thin-wall approximation is valid. 

This result has a very simple interpretation. CDL instantons describe tunneling through the barrier. Meanwhile, HM instantons in the inflationary regime can be  interpreted in terms of the Brownian motion, when the field slowly climbs to the top of the barrier due to accumulation of quantum fluctuations with the Hawking temperature $H/2\pi$. If the barrier is very wide, it is easier to climb the barrier rather that to tunnel through it. In this case the HM tunneling prevails and the stochastic/thermodynamic description of this process in terms of dS entropy is very useful. If the barrier is very thin, it is easier to tunnel, and CDL instantons are more efficient.

For the simple models with the parameters given in our paper one has $3T^2< 8V_1$, and the tunneling occurs mainly due to CDL instantons. Even in this case the stochastic/thermodynamic approach may remain useful: If this approach is valid not only in the context of inflationary cosmology with $|V''| \ll V$ but also in the situations with $|V''| \gtrsim V$, it provides a simple upper bound on the decay time of  dS space, $t_{\rm decay} < e^{\bf S(\varphi_0)-\bf S(\varphi_1)}$.

\section{Discussion}

It has been a difficult problem to construct realistic cosmologies from
string theory, as long as the moduli fields are not frozen.  In this note,
we have seen that it is possible to stabilize all moduli in a controlled manner
in the general setting of compactifications with flux.
This opens up a promising arena for the construction of string cosmologies.

More specifically, we have seen that it should be possible to construct
metastable dS vacua in the general framework of \cite{GKP}, by
including anti-branes \cite{KPV} and incorporating
non-perturbative corrections to the superpotential from
D3 instantons \cite{witsup} or low-energy gauge dynamics.   
For cosmology, it might be more interesting to include a more
complicated sector at the end of the warped throat.  For instance,
the D3/D7 inflationary models of \cite{DasKal}, or models with both branes and
anti-branes \cite{bantib}, could be of interest
in this context.  It should also be possible to make similar constructions
in the heterotic theory, perhaps using the models of \cite{bbdg} as
a starting point (and then incorporating a
mechanism to stabilize the dilaton
in a controlled manner).

It is worth summarizing why we believe the constructions in this paper
are reliable. Our analysis was carried out in the framework of supergravity.
There are two kinds of corrections to this analysis, in the $g_s$ and
$\alpha^{\prime}$  expansions.
By tuning the fluxes we have argued that  both of these can be controlled.
For example, it is reasonable to believe that $g_s$ can be stabilized at a value of order
$g_s \sim 0.1$ and the volume $\sigma \gg 1$ (for the volume this was discussed in  section II C).
Note that for reliability we do not
require that the dilaton and volume be made arbitrarily
small or big; in
fact this is not possible since the fluxes can only be tuned discretely.
We only require that these moduli take appropriately small or big values,
and this can be
achieved, especially if $M$ has enough three-cycles, yielding many
possible choices of flux background.

We were also able to prove, in section IV, that our dS minima
are short lived compared to the timescale for Poincar\'{e} recurrences.
This generalizes to any construction where the dS minimum is
separated from the Dine-Seiberg run-away vacuum (which is ubiquitous in
string theory) by a potential which remains non-negative; the
simplest controlled examples in string theory will have this feature. One
can also
imagine more complicated shapes of the potential between the dS
minimum and infinity, which include some intervening AdS critical
points. It is natural to wonder if a more
general statement can be obtained that would apply in these
cases as well.

Finally, it would be interesting to discover string theory
models which naturally incorporate both $> 60$ e-foldings
of early universe inflation, and a late-time cosmology in agreement
with the most recent data \cite{data}.
(To show that it is possible to obtain a small enough value of the
dS cosmological constant for late-time cosmology, one would have to
demonstrate that an idea along the lines of \cite{BP} can be
implemented in this context.)
Some further steps towards making more realistic cosmological toy-models
inspired by string theory, in the
same general framework as this note, will be presented
in \cite{toappear}.

It is a pleasure to thank I. Bena, A. Dabholkar,  M. Dine, S. Gukov, M. Headrick,  S. Hellerman,
L. Kofman, A. Maloney, A. Sen, E. Silverstein, A. Strominger,
L. Susskind, P. Tripathi,  and H. Verlinde for
useful discussions.  This work was supported in part
by NSF grant PHY-9870115. The work by A.L. was also supported by the Templeton Foundation grant
No. 938-COS273. The work of S.K. was also supported by
a David and Lucile Packard Foundation Fellowship for
Science and Engineering, NSF grant PHY-0097915 and
the DOE under contract
DE-AC03-76SF00515. S.P.T. acknowledges the support of the  DAE,
the Swarnajayanti Fellowship,  and most of all the people of India.


\begin{thebibliography}{4}

\bibitem{data}
S. Perlmutter et al. [Supernova Cosmology Project
Collaboration], ``Measurements of Omega and Lambda from 42 High-Redshift
Supernovae,'' Astrophys. J. {\bf 517}, 565 (1999) [astro-ph/9812133];
A. G. Riess et al. [Supernova Search Team Collaboration], ``Observational
Evidence from Supernovae for an Accelerating Universe and a
Cosmological Constant,'' Astron. J. {\bf 116}, 1009 (1998)
[astro-ph/9805201].

\bibitem{conceptual}
T. Banks, ``Cosmological Breaking of Supersymmetry?  Or Little Lambda
goes back to the future 2,'' [arXiv:hep-th/0007146];
S. Hellerman, N. Kaloper and L. Susskind, ``String theory and
quintessence,'' JHEP {\bf 0106}, 003 (2001) [arXiv:hep-th/0104180];
W. Fischler, A. Kashani-Poor, R. McNees and S. Paban, ``The
acceleration of the Universe: A challenge for string theory,''
[arXiv:hep-th/0104181];
E. Witten, ``Quantum Gravity in de Sitter Space,'' [arXiv:hep-th/0106109];
A. Strominger, ``The dS/CFT Correspondence,'' JHEP {\bf 0110}, 034 (2001)
[hep-th/0106113].

\bibitem{maldanun}
J. Maldacena and C. Nunez, ``Supergravity Description of Field
Theories on Curved Manifolds and a No Go Theorem,'' Int. J. Mod. Phys.
{\bf A16}, 822 (2001), [arXiv:hep-th/0007018];
G.W. Gibbons, ``Aspects of Supergravity Theories,'' in ${\it Supersymmetry,
Supergravity ~and~ Related~Topics}$, eds. F. del Aguila, J.A. de
Azcarraga and L.E. Ibanez (World Scientific 1985) pp.346-351;
B. de Wit, D.J. Smit and N.D. Hari Dass, ``Residual Supersymmetry
of Compactified $D=10$ Supergravity,'' Nucl. Phys. {\bf B283}, 165
(1987).






\bibitem{GKP}
S.~B.~Giddings, S.~Kachru and J.~Polchinski,
``Hierarchies from fluxes in string compactifications,''
Phys. Rev. {\bf D66}, 106006 (2002) [arXiv:hep-th/0105097].
%%CITATION = HEP-TH 0105097;%%


\bibitem{MSS}
E. Silverstein, ``(A)dS Backgrounds from Asymmetric Orientifolds,''
[arXiv:hep-th/0106209]; A. Maloney, E. Silverstein and A. Strominger,
``de Sitter Space in Noncritical String Theory,'' [arXiv:hep-th/0205316].


\bibitem{KPV}
S. Kachru, J. Pearson and H. Verlinde, ``Brane/Flux Annihilation and
the String Dual of a Non-Supersymmetric Field Theory,'' JHEP
{\bf 0206}, 021 (2002) [arXiv:hep-th/0112197].

\bibitem{dinesei}
M. Dine and N. Seiberg, ``Is the superstring weakly coupled?,''
Phys. Lett. {\bf B162}, 299 (1985).

\bibitem{Susskind}
L.~Dyson, J.~Lindesay and L.~Susskind,
``Is there really a de Sitter/CFT duality,''
JHEP {\bf 0208}, 045 (2002)
[arXiv:hep-th/0202163];
L.~Dyson, M.~Kleban and L.~Susskind,
``Disturbing implications of a cosmological constant,''
JHEP {\bf 0210}, 011 (2002)
[arXiv:hep-th/0208013];
N.~Goheer, M.~Kleban and L.~Susskind,
``The trouble with de Sitter space,''
arXiv:hep-th/0212209.

\bibitem{Acharya}
B.S. Acharya, ``A Moduli Fixing Mechanism in M-theory,''
[arXiv:hep-th/0212294].

\bibitem{john}
S. Hellerman, J. McGreevy and B. Williams, ``Geometric
Constructions of Non-geometric String Theories,'' [arXiv:hep-th/0208174];
A. Dabholkar and C. Hull, ``Duality Twists, Orbifolds and Fluxes,''
[arXiv:hep-th/0210209].

\bibitem{DasKal}
K. Dasgupta, C. Herdeiro, S. Hirano and R. Kallosh, ``D3/D7 Inflationary
Model and M-theory,'' Phys. Rev. {\bf D65}, 126002 (2002)
[arXiv:hep-th/0203019].

\bibitem{bantib}
S. Alexander, ``Inflation from D/Anti-D Brane Annihilation,''
Phys. Rev. {\bf D65}, 23507 (2002) [arXiv:hep-th/0105032];
G. Dvali, Q. Shafi and S. Solganik, ``D-Brane Inflation,''
[arXiv:hep-th/0105203];
C.P. Burgess, M. Majumdar, D. Nolte, F. Quevedo, G. Rajesh
and R. Zhang, ``The Inflationary Brane Anti-Brane Universe,''
JHEP {\bf 0107}, 047 (2001) [arXiv:hep-th/0105204].




\bibitem{KL}
R. Kallosh, A. Linde, S. Prokushkin and M. Shmakova,
``Supergravity, Dark Energy and the Fate of the Universe,''
[arXiv:hep-th/0208156].

\bibitem{toappear}
S. Kachru, R. Kallosh, A. Linde and S. P. Trivedi, in preparation.

\bibitem{Fre}
P. Fre, M. Trigiante and A. Van Proeyen, ``Stable de Sitter Vacua
from $N=2$ Supergravity,'' [arXiv:hep-th/0205119].

\bibitem{Panda}
M. de Roo, D.B. Westra and S. Panda, ``de Sitter solutions in $N=4$
matter coupled supergravity,'' [arXiv:hep-th/0212216].

\bibitem{Quevedo}
Y. Aghababaie, C.P. Burgess, S.L. Parameswaran and F. Quevedo,
``SUSY Breaking and Moduli Stabilization from Fluxes in Gauged
6d Supergravity,'' [arXiv:hep-th/0212091].


\bibitem{BP}
R. Bousso and J. Polchinski, ``Quantization of Four-Form Fluxes and
Dynamical Neutralization of the Cosmological Constant,''
JHEP {\bf 0006}, 006 (2000), [arXiv:hep-th/0004134].

\bibitem{Feng}
J. Feng, J. March-Russell, S. Sethi and F. Wilczek, ``Saltatory
Relaxation of the Cosmological Constant,'' Nucl. Phys. {\bf 602},
307 (2001) [arXiv:hep-th/0005276].

\bibitem{DRS}
K. Dasgupta, G. Rajesh and S. Sethi, ``M-theory, Orientifolds and
G-Flux,'' JHEP {\bf 9908}, 023 (1999), [arXiv:hep-th/9908088].

\bibitem{KST}
S. Kachru, M. Schulz and S. P. Trivedi, ``Moduli
Stabilization from Fluxes in a Simple
IIB Orientifold,'' [arXiv: hep-th/0201028].

\bibitem{PolFrey}
A. Frey and J. Polchinski, ``${\cal N}=3$ Warped Compactifications,''
Phys. Rev. {\bf D65}, 126009 (2002) [arXiv:hep-th/0201029].

\bibitem{Ferrara}
R. D'Auria, S. Ferrara and S. Vaula, ``$N=4$ gauged supergravity
and a IIB orientifold with fluxes,'' New J. Phys. {\bf 4}, 71
(2002) [arXiv:hep-th/0206241].

\bibitem{Frey}
A. Frey and A. Mazumdar, ``Three-Form Induced Potentials, Dilaton Stabilization,
and Running Moduli,'' [arXiv:hep-th/0210254].

\bibitem{Vafa}
C. Vafa, ``Evidence for F-theory,'' Nucl. Phys. {\bf B469}, 403
(1996) [arXiv:hep-th/9602022].

\bibitem{Sen}
A. Sen, ``Orientifold Limit of F-theory Vacua,'' Phys. Rev.
{\bf D55}, 7345 (1997) [arXiv:hep-th/9702165].

\bibitem{fourlist}
A. Klemm, B. Lian, S.S. Roan and S.T. Yau, ``Calabi-Yau Fourfolds
for M-theory and F-theory Compactifications,'' Nucl. Phys.
{\bf 518}, 515 (1998) [arXiv:hep-th/9701023].


\bibitem{GVW}
S. Gukov, C. Vafa and E. Witten, ``CFTs from Calabi-Yau Fourfolds,''
Nucl. Phys. {\bf B584}, 69 (2000) [arXiv:hep-th/9906070].

\bibitem{Taylor}
T. Taylor and C. Vafa, ``RR flux on Calabi-Yau and partial
supersymmetry breaking,'' Phys. Lett. {\bf B474}, 130 (2000)
[arXiv:hep-th/9912152].

\bibitem{Klemm}
G. Curio, A. Klemm, D. L\"ust and S. Theisen, ``On the Vacuum Structure
of Type II String Compactifications on Calabi-Yau Spaces with H Fluxes,''
Nucl. Phys. {\bf B609}, 3 (2001) [arXiv:hep-th/0012213].


\bibitem{noscale}
E. Cremmer, S. Ferrara, C. Kounnas and D.V. Nanonpoulos,
``Naturall vanishing cosmological constant in $N=1$ supergravity,''
Phys. Lett. {\bf B133}, 61 (1983); J. Ellis, A.B. Lahanas,
D.V. Nanopoulos and K. Tamvakis, ``No-scale Supersymmetric
Standard Model,'' Phys. Lett. {\bf B134}, 429 (1984).

\bibitem{Beckers}
K. Becker and M. Becker, ``M-theory on Eight Manifolds,'' Nucl.
Phys. {\bf B477}, 155 (1996) [arXiv:hep-th/9605053].

\bibitem{Verlinde}
H. Verlinde, ``Holography and Compactification,'' Nucl. Phys.
{\bf B580}, 264 (2000) [arXiv:hep-th/9906182];
C. Chan, P. Paul and H. Verlinde, ``A Note on Warped String Compactification,''
Nucl. Phys. {\bf B581}, 156 (2000) [arXiv:hep-th/0003236].

\bibitem{Mayr}
P. Mayr, ``On Supersymmetry Breaking in String Theory and its
Realization in Brane Worlds,'' Nucl. Phys. {\bf B593}, 99 (2001)
[arXiv:hep-th/9905221];
P. Mayr, ``Stringy Brane Worlds and Exponential Hierarchies,''
JHEP {\bf 0011}, 013 (2000) [arXiv:hep-th/0006204].


\bibitem{Greene}
B. Greene, K. Schalm and G. Shiu, ``Warped Compactifications in
M and F-theory,'' Nucl. Phys. {\bf B584}, 480 (2000) [arXiv:hep-th/0004103].

\bibitem{KS}
I.R. Klebanov and M.J. Strassler, ``Supergravity and a confining
gauge theory: duality cascades and $\chi$SB-resolution of naked
singularities,'' JHEP {\bf 0008}, 052 (2000) [arXiv:hep-th/0007191].



\bibitem{witsup}
E. Witten, ``Nonperturbative Superpotentials in String Theory,''
Nucl. Phys. {\bf B474}, 343 (1996) [arXiv:hep-th/9604030].

\bibitem{santrip}
P. Tripathy and S. P. Trivedi, ``Compactification with Flux on $K3$ and
Tori,'' [arXiv:hep-th/0301139].


\bibitem{kaplouis}
P. Candelas and H. Skarke, ``F-theory, $SO(32)$ and Toric Geometry,''
Phys. Lett. {\bf B413}, 63 (1997) [arXiv:hep-th/9706226];
V. Kaplunovsky and J. Louis, ``Phenomenological Aspects of F-theory,''
Phys. Lett. {\bf B417}, 45 (1998) [arXiv:hep-th/9708049].

\bibitem{bbhl}
K. Becker, M. Becker, M. Haack and J. Louis, ``Supersymmetry
Breaking and Alpha-Prime Corrections to Flux Induced Potentials,''
JHEP {\bf 0206}, 060 (2002), [arXiv:hep-th/0204254].

\bibitem{racetrack}
N.V. Krasnikov, ``On Supersymmetry Breaking in Superstring
Theories,'' Phys. Lett. {\bf B193}, 37 (1987);
L. Dixon, ``Supersymmetry Breaking in String Theory,''
SLAC-PUB-5229 (1990).


\bibitem{maldnast}
J. Maldacena and H. Nastase, ``The Supergravity Dual of a Theory
with Dynamical Supersymmetry Breaking,'' JHEP {\bf 0109}, 024 (2001)
[arXiv:hep-th/0105049].


\bibitem{Gibbons:1976ue}
G.~W.~Gibbons and S.~W.~Hawking,
``Cosmological Event Horizons, Thermodynamics, And Particle Creation,''
Phys.\ Rev.\ D {\bf 15}, 2738 (1977);
G.~W.~Gibbons and S.~W.~Hawking,
``Action Integrals And Partition Functions In Quantum Gravity,''
Phys.\ Rev.\ D {\bf 15}, 2752 (1977).



%%CITATION = HEP-TH 0212209;%%
%\cite{Coleman:1980aw}
\bibitem{Coleman:1980aw}
S.~R.~Coleman and F.~De Luccia,
``Gravitational Effects On And Of Vacuum Decay,''
Phys.\ Rev.\ D {\bf 21}, 3305 (1980).
%%CITATION = PHRVA,D21,3305;%%

%\cite{Hawking:1981fz}
\bibitem{Hawking:1981fz}
S.~W.~Hawking and I.~G.~Moss,
``Supercooled Phase Transitions In The Very Early Universe,''
Phys.\ Lett.\ B {\bf 110}, 35 (1982).
%%CITATION = PHLTA,B110,35;%%

\bibitem{GL} A.S. Goncharov and A.D. Linde, ``Tunneling in Expanding Universe: Euclidean and Hamiltonian Approaches,'' Sov. J. Part. Nucl.  {\bf 17},  369  (1986).

%\cite{Linde:1998gs}
\bibitem{Linde:1998gs}
A.~D.~Linde,
``Quantum creation of an open inflationary universe,''
Phys.\ Rev.\ D {\bf 58}, 083514 (1998)
[arXiv:gr-qc/9802038].
%%CITATION = GR-QC 9802038;%%


\bibitem{Bousso:1998vz}
R.~Bousso and A.~Chamblin,
``Patching up the no-boundary proposal with virtual Euclidean wormholes,''
Phys.\ Rev.\ D {\bf 59}, 084004 (1999)
[arXiv:gr-qc/9803047].

\bibitem{Bousso:1998ed}
R.~Bousso and A.~D.~Linde,
``Quantum creation of a universe with Omega not = 1: Singular and  non-singular instantons,''
Phys.\ Rev.\ D {\bf 58}, 083503 (1998)
[arXiv:gr-qc/9803068].

\bibitem{Banks:2002nm}
T.~Banks,
``Heretics of the false vacuum: Gravitational effects on and of vacuum
decay. II,''
arXiv:hep-th/0211160.

\bibitem{Gen:1999gi}
U.~Gen and M.~Sasaki,
``False vacuum decay with gravity in non-thin-wall limit,''
Phys.\ Rev.\ D {\bf 61}, 103508 (2000)
[arXiv:gr-qc/9912096].





\bibitem{Starobinsky:fx}
A.~A.~Starobinsky,
``Stochastic De Sitter (Inflationary) Stage In The Early Universe,'' in: {\it Current Topics in Field
Theory, Quantum Gravity and Strings}, Lecture Notes in Physics, eds.
H.J. de Vega and N. Sanchez (Springer, Heidelberg 1986) {\bf 206},
p. 107.




%\cite{Linde:nc}
\bibitem{book}
A.~D.~Linde,
``Particle Physics And Inflationary Cosmology,''  (Harwood, Chur, Switzerland, 1990).
%\href{http://www.slac.stanford.edu/spires/find/hep/www?irn=2352990}{SPIRES entry}


\bibitem{Hardart} A.~D.~Linde,
``Hard art of the universe creation (stochastic approach to tunneling and baby universe formation),''
Nucl.\ Phys.\ B {\bf 372}, 421 (1992)
[arXiv:hep-th/9110037].


\bibitem{LLM} A.~D.~Linde, D.~A.~Linde and A.~Mezhlumian,
``From the Big Bang theory to the theory of a stationary universe,''
Phys.\ Rev.\ D {\bf 49}, 1783 (1994)
[arXiv:gr-qc/9306035].


\bibitem{bbdg}
K. Becker, M. Becker, K. Dasgupta and P. Green, ``Compactifications
of Heterotic Theory on Non-K\"ahler Complex Manifolds: I,''
[arXiv:hep-th/0301161].



\end{thebibliography}
\end{document}